\documentclass[sigconf]{acmart}
\pdfoutput=1

\usepackage{multirow}
\usepackage{listings}
\def\BibTeX{{\rm B\kern-.05em{\sc i\kern-.025em b}\kern-.08em
    T\kern-.1667em\lower.7ex\hbox{E}\kern-.125emX}}

\AtBeginDocument{%
  \providecommand\BibTeX{{%
    \normalfont B\kern-0.5em{\scshape i\kern-0.25em b}\kern-0.8em\TeX}}}

\renewcommand\footnotetextcopyrightpermission[1]{} 
\makeatletter
\renewcommand\@formatdoi[1]{\ignorespaces}
\makeatother

\begin{document}

\title{Finding Security Vulnerabilities in Network Protocol Implementations}





\author{Kaled Alshmrany}
\affiliation{%
  \institution{University of Manchester}
  \city{Manchester}
  \state{UK}
}
\email{kaled.alshmrany@postgrad.manchester.ac.uk}

\author{Lucas C. Cordeiro}
\affiliation{%
  \institution{University of Manchester}
  \city{Manchester}
  \country{UK}
}
\email{lucas.cordeiro@manchester.ac.uk}

\begin{abstract}


Implementations of network protocols are often prone to vulnerabilities caused by developers' mistakes when accessing memory regions and dealing with arithmetic operations. Finding practical approaches for checking the security of network protocol implementations has proven to be a challenging problem. The main reason is that the protocol software state-space is too large to be explored. Here we propose a novel verification approach that combines fuzzing with symbolic execution to verify intricate properties in network protocol implementations. We use fuzzing for an initial exploration of the network protocol, while symbolic execution explores both the program paths and protocol states, which were uncovered by fuzzing. From this combination, we automatically generate high-coverage test input packets for a network protocol implementation. We surveyed various approaches based on fuzzing and symbolic execution to understand how these techniques can be effectively combined and then choose a suitable tool to develop further our model on top of it. In our preliminary evaluation, we used ESBMC, Map2Check, and KLEE as software verifiers and SPIKE as fuzzer to check their suitability to verify our network protocol implementations. Our experimental results show that ESBMC can be further developed within our verification framework called \textit{FuSeBMC}, to efficiently and effectively detect intricate security vulnerabilities in network protocol implementations. 
\end{abstract}

\maketitle
\setlength{\parskip}{.05cm plus4mm minus3mm}
\section{Introduction}

One of the most challenging and error-prone tasks is the implementation of network protocols. When implementing network protocols, software bugs introduced during their implementation can lead to devastating security vulnerabilities~\cite{vanhoef2018symbolic}. For example, a stack-based buffer overflow in Sami FTP Server 2.0.1 allowed remote attackers to execute arbitrary code via a long USER command, which triggered a buffer overflow when visualizing the log~\cite{Micro2011}. Vulnerabilities exist in network protocol implementations, such as DNS, TCP, and FTP~\cite{wen2016testing}. These vulnerabilities are hard to be detected by protocol formal verification methods since the protocol software state-space is too large to explore~\cite{wen2016testing}. The validation of possible events such as packet access, packet loss, and timeout must be checked during the protocol implementation~\cite{george2016protocol}.



There exist many reasons to verify network protocol implementations. First, the state-space exploration of protocol implementations is often large~\cite{george2016protocol}. Second, finding semantic errors need a machine-readable specification~\cite{george2016protocol} to check whether the implementation meets the specification automatically. Third, many bugs manifest themselves only after a long period of operation~\cite{george2016protocol}. Therefore, these problems have attracted the attention of many researchers to develop automated tools to check for errors in network protocols. However, to understand the problems of network protocol implementations and motivate research towards new techniques for detecting implementation errors, we need to consider real examples that show how ambiguities in protocol specifications can cause various flaws and interoperability problems during the protocol implementation. Network protocols are described in specifications that contain the information required to produce a protocol implementation. Protocol specifications are usually written in informal languages, and therefore may be vague, incomplete, or fail to address important properties of the protocol implementation. These specifications are referred to by multiple manufacturers and lead to different protocol implementations. A misinterpretation of a specification by one of the manufacturers can cause implementation errors. Many errors are not detected until the service is in real use. Therefore, there exists a strong need for developing a verification method to reduce such errors made by programmers, which cause security vulnerabilities in network protocol implementations.

Techniques such as fuzzing~\cite{miller1995fuzz}, symbolic execution~\cite{king1976symbolic}, static code analysis~\cite{faria2008inspections}, and taint tracking~\cite{qin2006lift} are the most common techniques to detect security vulnerabilities in network protocol implementations. Here we propose a novel method that combines fuzzing and symbolic execution to detect security vulnerabilities in network protocol implementations. We use two approaches for symbolically verifying network protocol implementations. One based on path exploration, i.e., when a branch is found, a symbolic executor explores each branch separately, thereby making a copy of the current state. Another one based on bounded model checking (BMC), i.e., BMC evaluates both branch sides and merges states after that branch. We also exploit fuzzing to produce random inputs to locate security vulnerabilities in network protocols.
On the one hand, fuzzing is generally unable to create various inputs that exercise all paths in the network protocol implementation. On the other hand, symbolic execution may also not achieve high path-coverage because of the state-space explosion problem. Consequently, fuzzing and symbolic execution by themselves often cannot reach the deep states of the network protocol implementation. As a consequence, the vulnerabilities related to the deep states cannot be identified and detected by these techniques. Therefore, a hybrid approach involving fuzzing and symbolic execution may achieve better function coverage than fuzzing or symbolic execution in isolation. 

%


Here we combine fuzzing and symbolic execution techniques in an unprecedented manner to exploit the process of detecting security vulnerabilities in network protocol implementations. In particular, our verification model requires enhancements to symbolic execution so that it can efficiently explore various network protocol implementations to avoid the path explosion problem. These enhancements need to provide both broad and deep exploration of the state-space, thus resulting in high source code coverage of the network protocol implementations. Our preliminary study provides a detailed survey and taxonomy on fuzzing and symbolic execution for verifying network protocol implementations. Thus, we make two significant contributions. First, we describe a novel verification technique that uses symbolic execution combined with fuzzing to generate automatically high-coverage test packets from the network protocol implementations. We detect various generic implementation errors, e.g., invalid memory access or division-by-zero, which can cause the implementation of the network protocol to crash. We use fuzzing for an initial exploration of the network protocol and then apply symbolic execution to generate high-coverage test packets for network protocol implementations automatically. Second, our experimental results show that ESBMC~\cite{esbmc2018} can be further developed to detect security vulnerabilities in network protocols. In particular, ESBMC was able to detect a security vulnerability targeted at the FTP implementation faster than other existing tools such as Map2check~\cite{menezes2018map2check}, KLEE~\cite{cadar2008klee}, and SPIKE~\cite{aitel2002advantages}.

\section{Preliminaries}
\label{sec:Background}

\subsection{Network Protocol}
\label{sec:NetworkProtocol}

A network can be defined as a group of entities that are interconnected with communication technologies and allow the exchange of information~\cite{A.Tanenbaum2002}. The communicating entities require an agreement for exchanging information, and these agreements are known as \textit{network protocols}. The messages exchanged by these entities are referred to as \textit{packets}. A sequence of packets is called a \textit{packet stream}. Information related to methods, behavior, and packet formats are described in documents when designing a network protocol \cite{wen2016testing}; these documents form the \textit{protocol specification}, which is routinely referenced by developers of a protocol implementation. Implementations of network protocols are referred to as \textit{network daemons} in UNIX and other operating systems. When requirements of a protocol $P$ are specified, these requirements are described in the protocol specification $S$, and the specification is implemented in $I$. For example, File Transfer Protocol (FTP) is a standard network protocol used for the transfer of computer files between a client and a server on a computer network \cite{wen2017protocol}; it is described in several Request For Comments (RFC) documents that form the protocol specification. Several implementations of the specification exist, such as FileZilla (Windows) and Pure-FTPd (Unix). FTP is an application-level protocol used on (TCP/IP) networks for file exchange. The FTP is implemented as follows. First, a client uses port 21 to connect to the server. The client requests are sent in ASCII using this socket. A new socket is opened on port 20 with the server when the client requests to transfer data. Client requests mostly consist of a four-letter message type followed by the actual message. The server responses are in ASCII, where the first three digits correspond to a status code followed by an optional message. We use the FTP implementation as our case study to evaluate our prototype.
\subsection{Fuzzing}

Fuzzing is a software testing technique to exploit vulnerabilities in software systems~\cite{munea2016network}. Fuzzing prepares random or semi-random inputs to the target network protocol. Critical security flaws most often occur because program inputs are not adequately checked~\cite{wang2013model}. Since these inputs are random, their unexpected and improper appearance in a target network protocol is highly probable. If the target network protocol does not reject these improper inputs, it will hang or crash during fuzz testing. Fuzzing is a quick and cost-effective method for locating security vulnerabilities in network protocols. Software systems that cannot endure fuzzing could potentially lead to security holes. For example, in the FTP protocol, a network analyzer such as Wireshark extracts specifications of a network protocol implementation from conversations recorded between server/client sessions. Once the network protocol specifications are extracted, the fuzzing engine is loaded with an extendable list of fuzzing functions. Initially, the fuzzing engine sets its state to the root of the protocol. It then monitors the input traffic, thereby making appropriate transitions and applying fuzzing functions.

\subsection{Symbolic Execution}

Symbolic execution is widely used to find security vulnerabilities by analyzing program behavior and generating test cases~\cite{chipounov2009selective}. Using symbolic input values instead of concrete input values is the primary concept of symbolic execution. This method treats the paths as symbolic constraints and solves the constraints to output a concrete input as a test case. In terms of network protocols, we extract the message formats from the protocol specification of the target network protocol implementation. Then, we use these message formats to construct a concrete packet, which is used to mark the ID field of this packet as symbolic values to form a symbolic packet. After that, we invoke the symbolic execution engine to explore possible paths of the protocol program using two approaches: path-based symbolic execution and BMC.  
%

\subsection{Challenges in Verifying Network Protocols}

There exist many challenges to verify network protocols. First, the network protocol size is large and complicated since it involves different communicating entities. Second, testing and verification is a long process in which many errors appear only after a long period of operation. Avoiding the path explosion is one of the most difficult challenges we may face when we use a symbolic execution engine. The possible number of execution paths considered is so large, which only a small part of the program state-space is explored. Lastly, if we test a protocol in a virtual environment is not the same as testing it in a real environment. Thus, detecting vulnerabilities in network protocols in the real-world is a challenge, which deserves specialized verification approaches to ensure behavior correctness.

\section{Finding Vulnerabilities in Network Protocol Implementations} 


Network protocol implementations are often prone to vulnerabilities caused by the developer mistakes, which include:
\textit{buffer overflow}, which is a situation where a running program attempts to write data outside the memory buffer, which is not intended to store this data~\cite{black2016defeating};
\textit{memory leak}, which occurs when programmers create a memory in a heap and forget to delete it~\cite{zhang2018novel};
\textit{denial-of-service attack} (DoS), which is a security event that occurs when an attacker prevents legitimate users from accessing specific computer systems, devices, services, or other IT resources~\cite{US-CERT2009}. 
As an example, a vulnerability in the Cisco Discovery Protocol (CDP) module of Cisco IOS XE Software Releases 16.6.1 and 16.6.2 could allow an unauthenticated, adjacent attacker to cause a memory leak, which could lead to a DoS condition~\cite{Cisco2018}.

\subsection{\textit{FuSeBMC} for detecting security vulnerabilities }




After searching for tools that combine fuzzing and symbolic execution, we have found just one promising tool: Driller~\cite{stephens2016driller}. There exist other methods that claim to combine fuzzing and symbolic execution, but we have not found any tool implementation available~\cite{zhang2017hybrid}. Given the current knowledge in software verification and testing, there exists no tool that has been developed in the field of network protocols, which require dealing with packets in the network is one of the challenges that were not addressed by tools that integrate the two technologies. However, some tools that do not combine the two technologies faced some problems such as path explosion or achieved low coverage. Therefore, we propose a novel approach called \textit{FuSeBMC} for detecting security vulnerabilities in network protocol implementations using fuzzing and symbolic execution. The main idea is to generate a set of test input packets using fuzzing to explore the state-space initially. These test inputs will guide the symbolic execution and BMC engines to explore the parts that fuzzing could not reach. In other words, it will legalize the scanning process and avoid problems such as code explosion. Then, use symbolic execution and BMC to achieve high-code coverage and replay them against an implementation, thereby observing potential violations of rules derived from the protocol specification. We devise an exploration method that achieves broad and in-depth exploration of the state-space of a target implementation.

\begin{figure}
    \includegraphics[width=1.0\columnwidth]{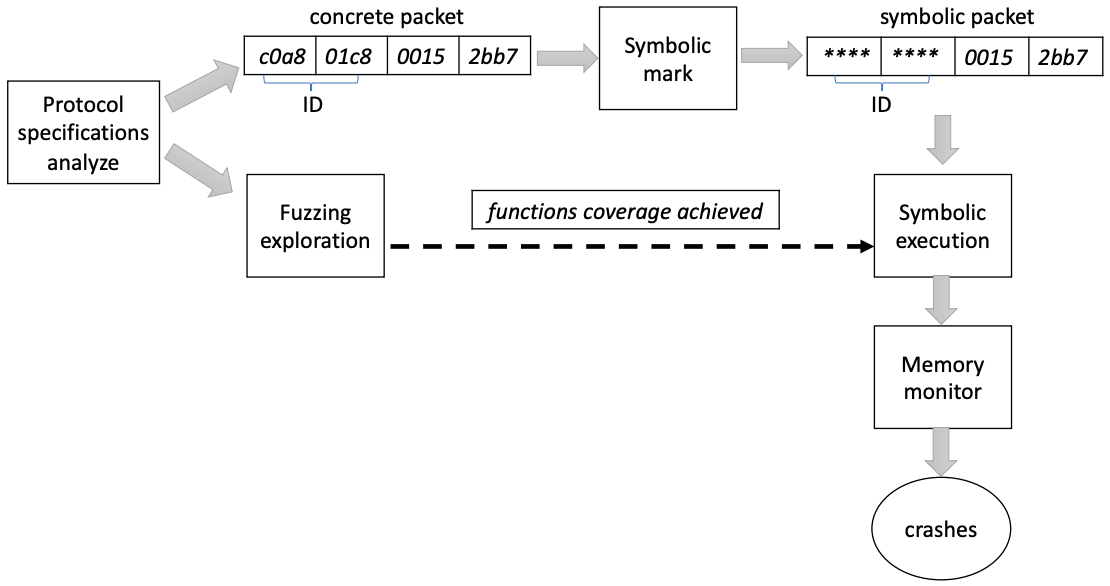}
    \caption{\textit{FuSeBMC} Verification Framework.}
    \label{fig:framework}
\vspace{-4ex}
\end{figure}

Our prototype is illustrated in Fig.~\ref{fig:framework}. There exist five steps to verify network protocol implementation. First, the protocol specification analyzer produces the concrete packet. We use here the Wireshark~\cite{Combs} to capture the packet that will be sent between the client and the FTP server. Second, our prototype employs American Fuzzy Lop (AFL)~\cite{Zalewski2015}, which is a tool that uses initial test-cases and genetic mutations to generate new test cases. AFL works with the target software that accepts inputs from the standard input or a file. We fuzz the software for exploring the functions. Then, we compute the function coverage achieved by the fuzzer. Third, we mark the input packet as a symbolic packet that would result in too many paths, and most of these paths would not increase code coverage because they would refer to invalid packets, which are usually discarded by an implementation. Typically a network packet consists of multiple fields, which are part of the packet header. Most protocol implementations contain logic for handling these fields. Therefore, \textit{FuSeBMC} uses these fields as symbolic variables instead of entire input packets. Fourth, we use path-based symbolic execution and BMC to reach those functions that were uncovered by the fuzzer. Because of an initial exploratory phase with fuzzer, a symbolic execution engine becomes capable of excluding the functions that can be reached quickly and, hence, more time can be allocated to intricate and deep functions. Lastly, the symbolic marker converts the concrete packet to a symbolic packet by marking some bytes of the packet as symbolic values.

Our \textit{FuSeBMC} prototype builds on top of Map2Check~\cite{menezes2018map2check} as a path-based symbolic execution engine combined with fuzzing and ESBMC~\cite{esbmc2018} as a state-of-the-art BMC engine. Both tools explore all program paths of the network protocol software and produce a concrete packet while the memory monitor module reports and records the crashes. For example, if the user provides an instruction to mark the flags field as symbolic, our prototype replaces the concrete value of this field within the packet with symbolic values while keeping the other fields concrete. Thus, \textit{FuSeBMC} explores possible execution paths corresponding to the various input packets having different flags values. At the end of each execution path, it stores the concrete test packet for a given path on disk.








\section{Experimental Evaluation}

\subsection{Benchmarks and Setup}


When we designed \textit{FuSeBMC}, we defined two main criteria to hold in our prototype. First, the ability to detect bugs that can be evaluated by validating network protocol implementations against their protocol specifications. Second, the low verification time to find security vulnerabilities when compared to existing approaches.
We conducted experiments on a simple File Transfer Protocol (FTP) written in the C programming language and a server (Vulnserver), also written in C that is known to have a security vulnerability. Although these two C programs are simple implementations of the FTP protocol, they allow us to evaluate the main elements of our proposed verification method initially. Additionally, in order to compare the effects of vulnerability detection, we compared our method with SPIKE~\cite{aitel2002advantages}, which is based on generational fuzzer for protocol implementations. In our experiments, we have considered two sets of properties: ``buffer overflow'' and ``arithmetic overflow''. The experiments were conducted by hosting two machines, namely, a \textit{host} machine and a \textit{guest} machine. The host machine included an Intel Core i7 2.9 GHz CPU with 16 GB memory and was running Ubuntu 14.04. The guest machine was running Windows 10. The tools were running Ubuntu 14.04 system and the fuzzing tool SPIKE was running Kali Linux on the host machine as a virtual system.
All benchmarks, tools, and results for this evaluation are available on a supplementary web page.\footnote{\url{https://drive.google.com/drive/folders/1pD7UW4JWtfKcAXzd2ea9Ma5jL_lJwpED}} 

\subsection{Objectives}
The main goal of this evaluation is to check the performance and suitability of these tools ESBMC~\cite{gadelha2019esbmc}, KLEE~\cite{cadar2008klee}, and Map2Check~\cite{menezes2018map2check} to be further developed to detect security vulnerabilities in real-world network protocol implementation within \textit{FuSeBMC}. Our experimental evaluation aims to answer two experimental goals:

\begin{enumerate}
\item[EG1] \textbf{(vulnerability detection)} Are ESBMC, KLEE and Map2Check able to detect security vulnerabilities in real-world network protocol implementations?
\item[EG2] \textbf{(witness validation)} Are ESBMC, KLEE and Map2Check able to provide further evidence of the detected security vulnerabilities?
\end{enumerate}

\subsection{Results}
 
We applied ESBMC, KLEE, and Map2Check to the simple FTP server implementations. Table~\ref{table:results} shows our experimental results. ESBMC has found an array-bounds violation, KLEE could not find any property violation, while Map2check provided the result ``UNKNOWN''. These results partially answer \textbf{EG1}, given that only ESBMC can detect security vulnerabilities in network protocol implementations. Additional implementation effort would be needed to adapt KLEE and Map2Check to verify protocol implementations successfully.
\begin{table}[]
\scalebox{0.8}{%
\begin{tabular}{|l|l|l|l|l|l|}
\hline
\textbf{Benchmark}          & \textbf{LOC}         & \textbf{Tool} & \textbf{Result} & \textbf{Vulnerability} & \textbf{Time} \\ \hline
\multirow{3}{*}{FTP Server} & \multirow{3}{*}{387} & ESBMC         & FALSE           & Buffer Overflow        & \textbf{\textless{}1s} \\ \cline{3-6} 
                            &                      & KLEE          & TRUE         & Not detected           & 2s            \\ \cline{3-6} 
                            &                      & Map2Check     & UNKNOWN         & Not detected           & 2s            \\ \hline
\multirow{3}{*}{Vulnserver} & \multirow{3}{*}{248} & ESBMC         & FALSE           & Buffer Overflow        & \textbf{\textless{}1s} \\ \cline{3-6} 
                            &                      & SPIKE         & Crashed         & Buffer Overflow        & 8s            \\ \cline{3-6} 
                            &                       & KLEE          & FALSE          & Buffer Overflow        & 2s            \\ \cline{3-6}
                            &                      & Map2Check     & UNKNOWN         & Not detected           & \textless{}1s \\ \hline
\end{tabular}}
\caption{Results of the FTP server and Vulnserver experiments.}
\vspace{-5ex}
  \label{table:results}
\end{table}

ESBMC, KLEE, Map2Check, and SPIKE were also executed on the server (Vulnserver), which has a known ``buffer overflow'' vulnerability. Table~\ref{table:results} shows our experimental results. ESBMC, SPIKE, and KLEE have found the ``buffer overflow'' vulnerability, while Map2Check could not detect this vulnerability. Additionally, we compared these tools in terms of verification time; we have found that ESBMC can detect the ``buffer overflow'' vulnerability in less than one second, while SPIKE took about $8$s and KLEE took $2$s to find that vulnerability. We have observed that ESBMC provides a counterexample as a sequence of stages from the initial to the bad state to reproduce the vulnerability, which answers \textbf{EG2}.
%



\subsection{Threats to Validity}

The main threat to the validity of our experiments is that we performed our evaluation on a simple FTP protocol, to be able to verify the real-world FTP protocol. However, this assumption may not make our experiments in the simple FTP protocol work appropriately in the real-world FTP protocol. Additionally, the diversity of network protocols in terms of protocol specifications and packets format also affects the accuracy of our experiments. As a result, the verification process might encounter a great challenge to deal with these differences in the protocols.

\section{Limitations, Related Work and Future Directions}

For more than $20$ years, network protocol implementation vulnerabilities have been mainly identified by fuzzing~\cite{barton1990fault}. Numerous instruments do exist, which aimed at network protocol implementations. SPIKE~\cite{aitel2002advantages} is a framework that provides an API to aid in creating fuzzed network protocol implementations. Additionally, PROTOS~\cite{kaksonen2001software} produces input parcels shrewdly based on protocol specifications. Nonetheless, these techniques cannot reach the deep states of network protocols and do not shape stateful protocols~\cite{wen2017model}. 

Symbolic execution has been used to identify security vulnerabilities and test network protocol. SymbexNet~\cite{song2014symbexnet} relied on a symbolic engine tool named KLEE~\cite{cadar2008klee} to combine symbolic execution with rule-based specifications. SymNet~\cite{sasnauskas2012integration} has been used to test the unmodified protocol implementations running on various operating systems. This SymNet works on the QEMU~\cite{bellard2005qemu} virtual machine, and it relies on KLEE as its symbolic engine. Furthermore, SymNet was designed on top of the S2E~\cite{chipounov2011s2e}, which is a platform for analyzing the properties and behavior of software systems.

The combination of symbolic execution and fuzzing has been proposed before. Driller~\cite{stephens2016driller} is a hybrid vulnerability excavation tool, which leverages fuzzing and selective concolic execution in a complementary manner, to find bugs deeply. The authors could avoid the path explosion inherent in concolic analysis and the incompleteness of fuzzing by combining the strengths of the two techniques and mitigate the weaknesses. Driller uses selective concolic execution to explore the paths that the Fuzzer engine can easily reach. However, when the fuzzer engine stuck, Driller generates symbolic input to explore the remaining paths.


Traditional fuzzing and symbolic execution techniques do not make full utilization of the protocol state information if they work independently; because of that, those two techniques have difficulties in reaching the deep states of network protocol implementations. As a result, the vulnerabilities related to deep states cannot be identified and detected by these existing methods. Therefore, a hybrid technique involving fuzzing and symbolic execution may achieve better function coverage than fuzzing or symbolic execution in isolation. This paper proposes an approach that helps reduce gaps in protocol implantation by dealing with network packets. In particular, we inject symbolic packets into the network so that one packet can generate various packets that help us test the target protocol, which makes our approach novel if compared to other existing approaches such as Driller~\cite{stephens2016driller}. In the future, we will develop a fully automated tool based on AFL and ESBMC for detecting security vulnerabilities in network protocol implementations by making full utilization of the protocol state information. 

Besides, we will use an FSM (Finite-State Machine) model to guide the symbolic execution. In detail, we will apply the L~\cite{angluin1987learning} online learning algorithm to construct the FSM model. We will infer the FSM model for the network protocol implementation by leveraging the L inference algorithm. This approach aims to learn the state machine by sending network protocol packets and observing the response packets. Thus, we will handle other vulnerabilities such as packet loss, timeout, and other semantic errors since the FSM allows one state active at a time.




\bibliographystyle{ACM-Reference-Format}
\bibliography{References.bib}

\end{document}